\documentclass[a4paper,fleqn,usenatbib,useAMS]{mnras}
\date{\today}
\usepackage{graphicx}
\usepackage{amsmath}

\newcommand{\bvec}[1]{\boldsymbol{#1}}
\def\ps{pseudo-caustic}
\def\pss{pseudo-caustics}
\def\Msun{{\rm M_\odot}}
\def\kpc{{\rm kpc}}

\def\kms{{\rm km\, s^{-1}}}
\def\actaa{Acta Astronomica}
\usepackage{verbatim}
\newcommand{\bvr}{\boldsymbol{r}}
\newcommand\be            {\begin{equation}}
\newcommand\ee            {\end{equation}}
\newcommand\ba            {\begin{aligned}}
\newcommand\ea            {\end{aligned}}
\usepackage{ae,aecompl}
\usepackage{txfonts}

\title{Detecting Extrasolar Asteroid Belts Through Their Microlensing Signatures}

\author[E. Lake, Z. Zheng, and S. Dong]{Ethan Lake$^1$\thanks{Contact e-mail: \href{lake@physics.utah.edu}{lake@physics.utah.edu}}, Zheng Zheng$^1$, and Subo Dong$^2$\\
$^1$Department of Physics and Astronomy, University of Utah,
    115 South 1400 East, Salt Lake City, UT 84112, USA\\
$^2$Kavli Institute for Astronomy and Astrophysics, Peking University, 
Yi He Yuan Road 5, Hai Dian District, Beijing 100871, China}

\date{\today}
\pubyear{2016}

\begin{document}
\label{firstpage}
\pagerange{\pageref{firstpage}--\pageref{lastpage}}
\maketitle

\begin{abstract}
We propose that extrasolar asteroid belts can be detected through their 
gravitational microlensing signatures. Asteroid belt + star lens systems 
create so-called ``pseudo-caustics'', regions in the source plane where the 
magnification exhibits a finite but discontinuous jump. 
%These features allow 
%such systems to generate distinctive microlensing light curves across a wide 
%region of belt parameter space and %possess remarkably large lensing cross-sections. 
These features allow such systems to generate distinctive signatures in the
microlensing light curves for a wide range of belt configurations, with
source trajectories as far as tenths of the Einstein ring radius from the 
centre of the lens.
%{ extend radially outward from the lens, creating a detectable light curve signature 
%as far as $0.3$ Einstein radii from the lens. }
Sample light curves for a range of asteroid belt 
parameters are presented. In the near future, space-based microlensing 
surveys like {\it WFIRST}, which will have the power of
detecting percent-level changes in microlensing light curves even with 
sub-minute exposure times, 
may be able to discover extrasolar asteroid belts with masses of the order of an
earth mass.

\end{abstract}


\section{Introduction}

Gravitational microlensing has established itself as a valuable tool for 
detecting objects covering a wide range of mass \citep[e.g.,][]{Paczynski86}.
Successful microlensing surveys began over two decades ago with target star 
fields in both the Galactic bulge \citep{Udalski93} and the Small
and Large Magellanic Clouds (SMC and LMC; \citealt{Alcock93,Aubourg93}). 
As pointed out by \citet{Mao91}, microlensing
can be used to probe extrasolar planetary systems. Following the discovery
of exoplanets through the microlensing technique \citep[e.g.,][]{Bond04,
Udalski05,Beaulieu06}, the number of detections has been steadily increasing
into the several dozens \citep[e.g.,][]{Kains13, Gould14, Suzuki14, Batista14}.
Owing to its sensitivity to low-mass planets and planets beyond the snow line,
microlensing has emerged as an important method to detect and study exoplanets
\citep[e.g.,][]{Gaudi10}. In addition, microlensing can also be applied to 
detect and characterize  free-floating planets \citep{Gaudi02,Sumi11}. 

It has also been suggested that microlensing can lend itself to the 
observational study of various parts of the planet-formation process, like 
circumstellar 
discs around source stars \citep{Zheng05} and lens stars \citep{Hundertmark09}.
In this paper, we propose to use microlensing to detect extrasolar asteroid 
belts around lens stars.

The observational study of extrasolar asteroid belts and debris discs has 
recently become an active field of study. Asteroid-belt and Kuiper-belt like 
analogs have been inferred around several extrasolar systems from the observed 
infrared excess of their host stars \citep[e.g.,][]{Chen01,Beichman05,
Moerchen07,Chen09,Backman09,Moerchen10}, resulting from dust generated by 
collisions of asteroids. Multiple asteroid-sized objects have also been
found to transit a white dwarf \citep[e.g.,][]{Vanderburg15,Croll15}. Residual
timing variations in millisecond pulsars are proposed as originating from 
asteroid belts \citep[e.g.,][]{Shannon13}. 
% All the same, the difficulty of these observations means that it is not well 
% known how common asteroid belts really are -- for example, our solar 
% system's asteroid belt is well below the current observational detection 
% threshold, and observations can only detect belts that are very spatially 
% extended. 
Determining the prevalence of extrasolar asteroid belts could shed insight 
into the planet formation process and even help estimate the frequency of 
extraterrestrial life \citep{Martin13}. Therefore, it would be valuable if 
we have complementary and effective methods to probe extrasolar asteroid belts
around main sequence stars.

In this paper, we show that microlensing by the lens system composed of a 
star and an asteroid belt can be a promising detection method. Microlensing is useful 
in this regard because of its sensitivity to low-mass objects located near 
the Einstein ring radius, which is typically on the order of a few AU. We find 
that asteroid belt + star lenses produce a wide variety of so-called 
``pseudo-caustics'', loci in the source plane across which the 
magnification of a source has a finite jump 
\citep[e.g.,][]{Evans98,Rhie10,Lee14,Lake16}. A detailed investigation
on the magnification properties for the general ring/belt+point lens has 
been presented in \citet{Lake16}.
In this paper, we show that sources passing over the pseudo-caustics can 
leave unique signatures in the resulting microlensing light curves, which
should be easily identifiable 
in upcoming observational surveys. 
{ Furthermore, the \pss\ have a 
comparatively large radial extent in the source plane, meaning that the 
probability for source trajectories to cross \pss\ can be significantly higher 
than to cross the formal caustics associated with planetary 
microlensing events.}

%\begin{figure*}
%\label{fig:coords}
%\plotone{coordinate_setup.pdf}
%\caption{An illustration of the geometry of our lensing system. Rays emitted from the star $S$ at an angle $\bvr_S$ in the source plane pass through the image plane at an angular position $\bvr_I$ and are deflected by the lensing system $L$ to an observer $O$. We let $a$ be the semi-major axis of the ring, so that the ring's projected semi-axes in the lens plane are $a$ and $b = a\cos i$.}
%\end{figure*}

The structure of this paper is as follows. In Section~\ref{sec:pss}, we recall key gravitational lensing equations and describe the \pss\ that are formed by the lenses we consider. Section~\ref{sec:lcurves} is devoted to the presentation of model light curves for a few illustrative cases. Finally, Section~\ref{sec:dis} consists of a summary and a discussion of our results. 

\section{Magnification properties of belt+star lenses}\label{sec:pss}

\subsection{Lensing equations}

The lens systems we consider are composed of a star and an asteroid belt.
In each model, 
the circularly symmetric belt is centred at the star. We denote the 
distance between the observer and the lens (source) as $D_L$ 
($D_S$) and the distance between the lens and source as $D_{LS}$ 
($D_S=D_L+D_{LS}$). 
The belt is described by an inner radius $a_i$ and an outer
radius $a_o$, the inclination angle $i$
(i.e., the angle between the normal of the belt plane and the line of 
sight direction), and the mass ratio $q = M_{\rm belt} / M_{\rm star}$ of the belt to the central lens 
star. We assume the asteroid belt has a constant mass density. This is obviously an idealisation. For example, the mass distribution of
asteroids and the discreteness effect \citep[e.g.][]{Heng09} are not accounted
for. However, it allows us to gain a clear understanding for how asteroid
belts as a whole can be expected to behave as lenses.

We express the angular position vectors in the image (lens) plane and 
source plane as $\bvr_I = (x_I,y_I)$ and $\bvr_S = (x_S,y_S)$, respectively, 
with the lens star located at the origin and the $x$-axis oriented
along the major axis of the projected ring/belt. Throughout, we normalize all angular vectors and ring/belt 
sizes by the Einstein ring radius of the combined ring/belt+star system,
\begin{equation} \label{eq:er}
\theta_E = \sqrt{\frac{4GM}{c^2}\frac{D_{LS}}{D_LD_S}},
\end{equation}
where $M = M_{\rm belt} + M_{\rm star}$ is the total mass of the lensing 
system. In the lens plane, the physical scale corresponding to $\theta_E$ is approximated by 
\begin{equation}\label{eq:xi0}
D_L \theta_E \approx 4 \left(\frac{D}{2\kpc}\right)^{1/2} \left(\frac{M}{\rm 1M_\odot}\right)^{1/2} {\rm AU},
\end{equation}
where $D$ satisfies $1/D = 1/D_L + 1/D_{LS}$.

The general lens equation is
\begin{equation}
\bvec{r}_S=\bvec{r}_I-\bvec{\alpha}(\bvr_I),
\end{equation}
where $\bvec{\alpha}(\bvr_I)$ is the normalized deflection angle caused by the lensing system for light rays at $\bvr_I$ in the image plane.

The deflection angle has contributions from both the star and the 
belt: $\bvec{\alpha}(\bvr_I) = \bvec{\alpha}_{\rm star}(\bvr_I) + \bvec{\alpha}_{\rm belt}(\bvr_I)$. For a belt with belt-to-star mass ratio $q$, inclination $i$, and inner and outer semi-major axes $a_i$ and $a_o$ respectively, 
we find \citep[][]{Schramm90,Lake16}
\be 
 \bvec{\alpha}_{\rm star}(\bvr_I) = \frac{1}{1+q}\frac{\bvr_I}{r_I^2}
\ee
and
\be
\bvec{\alpha}_{\rm belt}(\bvr_I) = \frac{q}{1+q}\frac{2}{a_o^2 - a_i^2} 
\left[ \frac{\tilde{a}_o^2}{a_o'+b_o'} 
\left(\frac{\bvec{x}_I}{a_o'} + \frac{\bvec{y}_I}{b_o'}\right) 
- \frac{a_i^2}{a_i'+b_i'} 
\left(\frac{\bvec{x}_I}{a_i'} + \frac{\bvec{y}_I}{b_i'}\right)\right].
\label{eqn:deflect_belt}
\ee
The primed semi-axes are defined by $a_i'^2 = a_i^2 + \lambda_i$ and 
$b_i'^2 = a_i^2\cos^2i + \lambda_i$, with $\lambda_i$ found by solving 
\be 
\frac{x_I^2}{a_i^2 + \lambda_i} + \frac{y_I^2}{a_i^2\cos^2i + \lambda_i} = 1,
\ee
which is the inner edge's confocal ellipse passing through $(x_I, y_I)$.
For $\tilde{a}_o$, $a'_o$ and $b'_o$, the relation is similar,  
$a_o'^2 = \tilde{a}_o^2 + \lambda_o$ and $b_o'^2 = \tilde{a}_o^2\cos^2i + 
\lambda_o$, and $\lambda_o$ is defined by
\be 
\frac{x_I^2}{\tilde{a}_o^2 + \lambda_o} + \frac{y_I^2}{\tilde{a}_o^2\cos^2i + \lambda_o} = 1.
\label{eqn:tildeao}
\ee
Depending on the location of the point $(x_I, y_I)$, we have three
regimes. First, if $(x_I, y_I)$ is inside the inner edge of the belt [i.e.
$x_I^2/a_i^2+y_I^2/(a_i\cos i)^2<1$], we set $\tilde{a}_o=a_i$ and the 
deflection angle by the belt [eq.~(\ref{eqn:deflect_belt})] becomes zero. 
Second, if $(x_I, y_I)$ is within the 
belt, we set $\lambda_o=0$ and $\tilde{a}_o$ is found from 
equation~(\ref{eqn:tildeao}), which is the semi-major axis of the ellipse 
centered at the origin and passing through $(x_I, y_I)$ with axis ratio 
of $\cos i$. Third, if $(x_I, y_I)$ is outside the outer edge of the belt
[i.e. $x_I^2/a_o^2+y_I^2/(a_o\cos i)^2>1$], we set $\tilde{a}_o=a_o$ and 
$\lambda_o$ is solved from equation~(\ref{eqn:tildeao}), and $a'_o$ and $b'_o$ 
have the meaning of the semi-major axes of the outer edge's confocal ellipse 
passing through $(x_I, y_I)$.

\citet{Lake16} perform a study on the gravitational lensing
properties with a system made of a point mass surrounded by a ring/belt
and focus their discussions on examples with large $q$. We refer the readers to
the above paper for details. In this paper, we present cases with low 
mass ratios that are more suitable for the study of extrasolar asteroid belts. 
%We employ the inverse ray shooting technique to obtain the magnification
%map of the lensing system, which forms the basis of our analysis.

\begin{figure*}
\includegraphics[scale=.7]{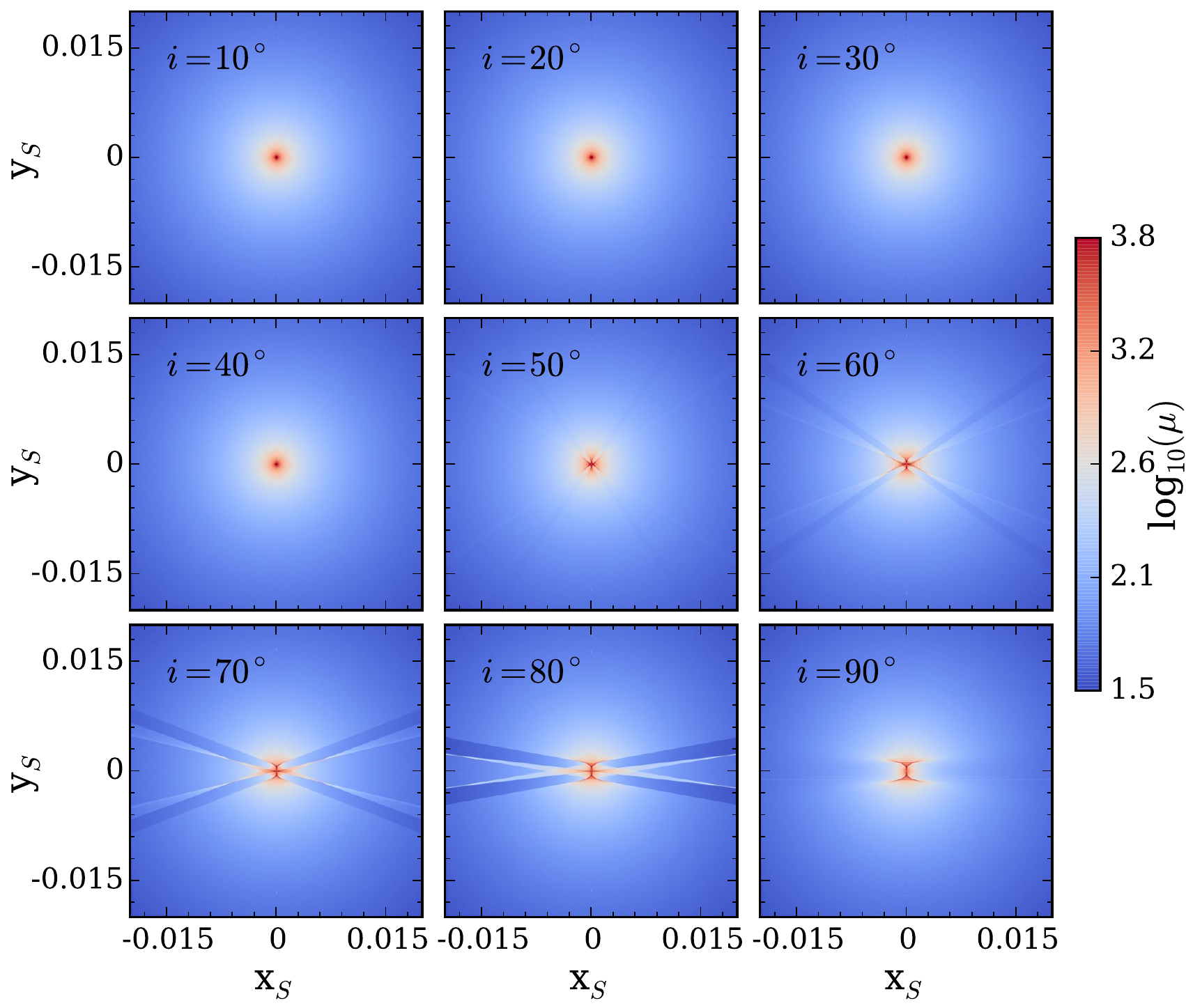}
\caption{
\label{fig:panel_mag}
Magnification maps near the centre region in the source plane for a 
lens composed of a central star
and a belt with inner and outer semi-major axes $a_i = 1.2$ and $a_o = 1.4$, 
a belt-to-stellar mass ratio of $q = 10^{-3}$ 
and plotted across various inclinations. For low inclinations the belt acts as a point mass, and the perturbations it induces to the star's magnification map are insignificant. As the inclination passes $\sim 50^{\circ}$, an X-shaped \ps\ feature appears. As the inclination increases the limbs of the X fold down and become wider, eventually merging when the ring is seen edge-on. A source moving across these features generates a very distinctive light curve (see text for more details). The features seen in this plot remain qualitatively unchanged for smaller $q$.}
\end{figure*}

\begin{figure}
\includegraphics[width=\columnwidth]{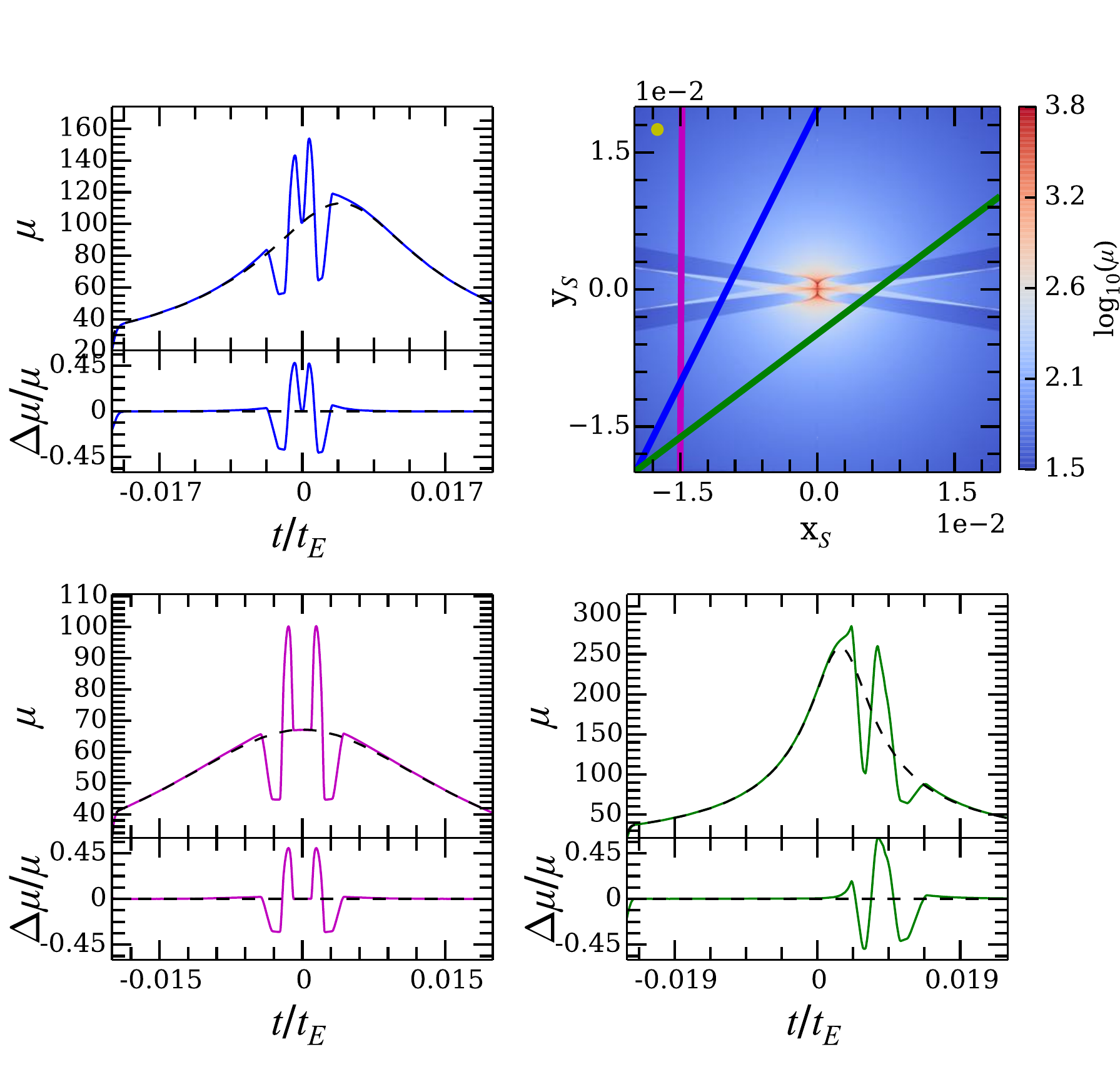}
\caption{\label{fig:a12to14q1e-3_lcurves}
Sample microlensing light curves for a lens with a central star and a belt with semi-major axes  
$a_i = 1.2$, $a_o = 1.4$, mass ratio $q=10^{-3}$, and inclination 
$i = 80^{\circ}$. The top-right panel shows the magnification near the centre of
the source plane. The circle at the top-left corner represents the dimensionless size of the source star. The other three panels show light curves corresponding to the given trajectories in the source plane, with the black dashed lines showing the light curve for the star-only case. The bottom subpanels show the relative deviation between the star+belt and star-only light curves. The source is
assumed to move along each trajectory in the top-left panel from bottom-left to
top-right, with $t = 0$ set to be at the halfway point of the trajectory.
%As the time $t / t_E$ increases on the $x$-axis, the source moves along the trajectory in the top left panel from bottom left to top right, with $t = 0$ set at the halfway point of its trajectory. %The maximum relative magnification caused by the \pss is nearly constant at $\sim50\%$. 
}
\end{figure}

\begin{figure}
\includegraphics[width=\columnwidth]{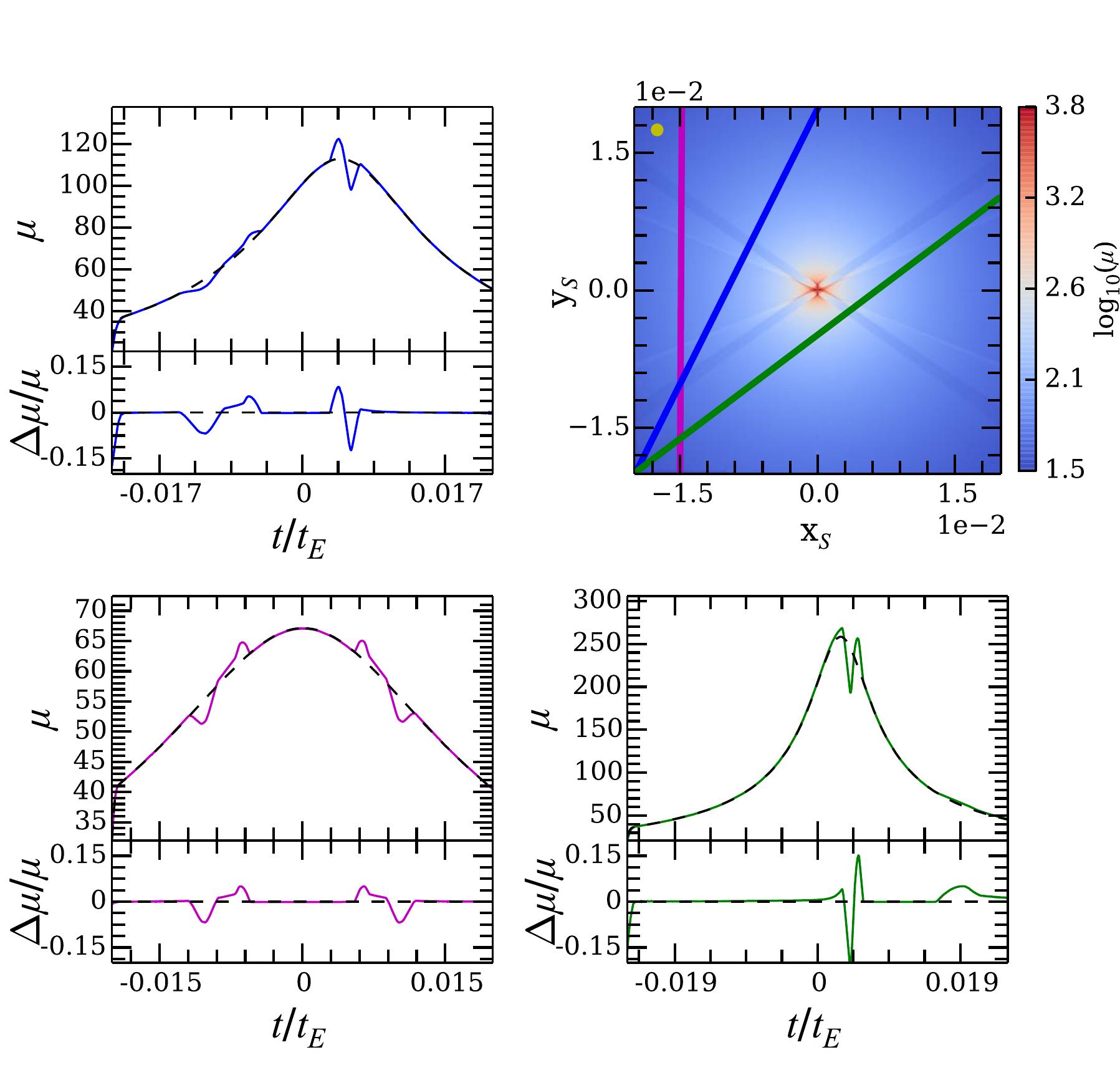}
\caption{ \label{fig:a12to14_60_q1e-3_lcurves}
Same as in Figure~\ref{fig:a12to14q1e-3_lcurves}, but with the inclination of the belt changed to $i = 60^{\circ}$. 
}
\end{figure}

\begin{figure}
\includegraphics[width=\columnwidth]{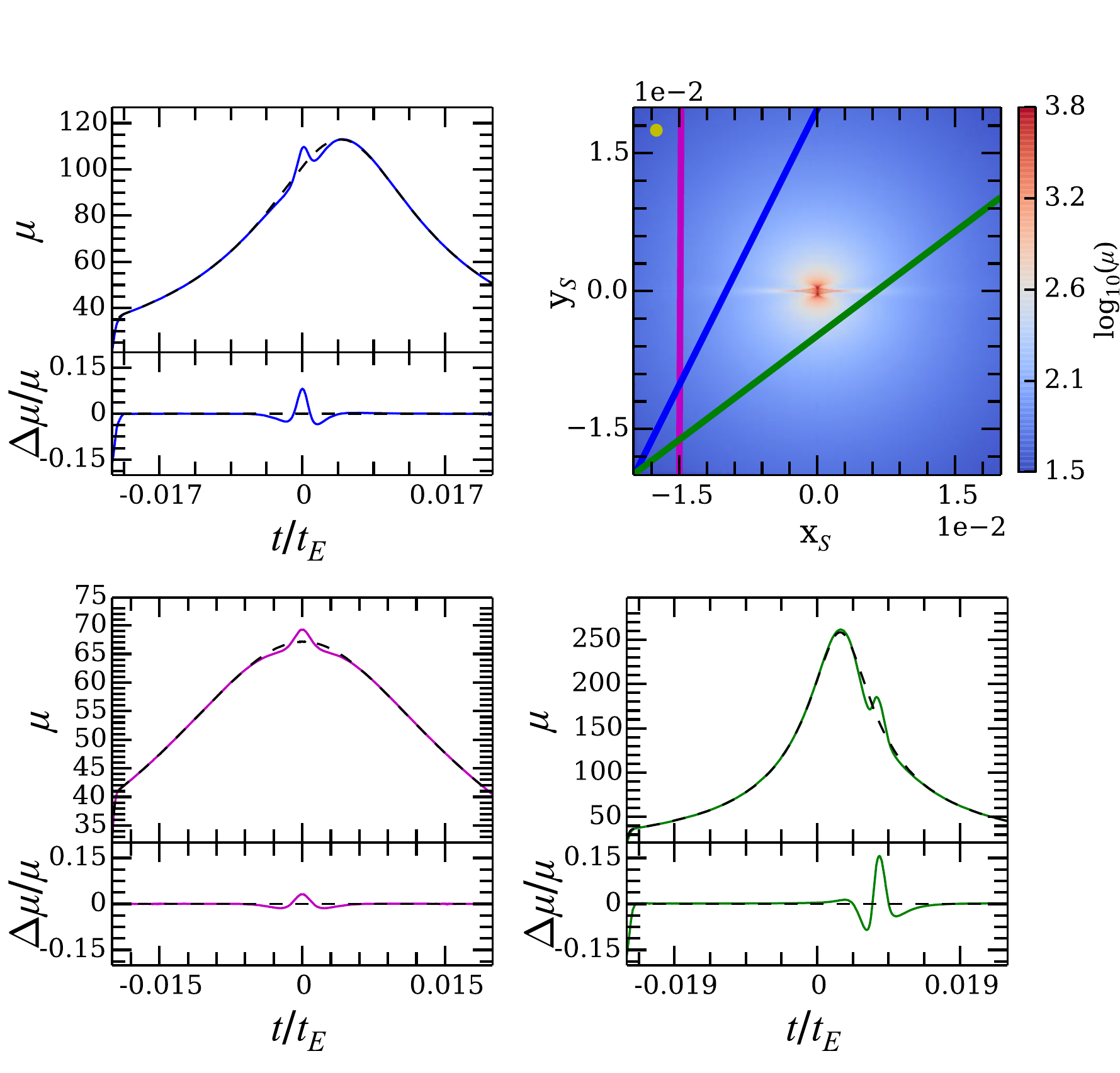}
\caption{\label{fig:a07to09_q1e-3_lcurves}
Same as in Figure~\ref{fig:a12to14q1e-3_lcurves}, but with semi-major axes $a_i = 0.7$ and $a_o = 0.9$.
}
\end{figure}

\begin{figure}
\includegraphics[width=\columnwidth]{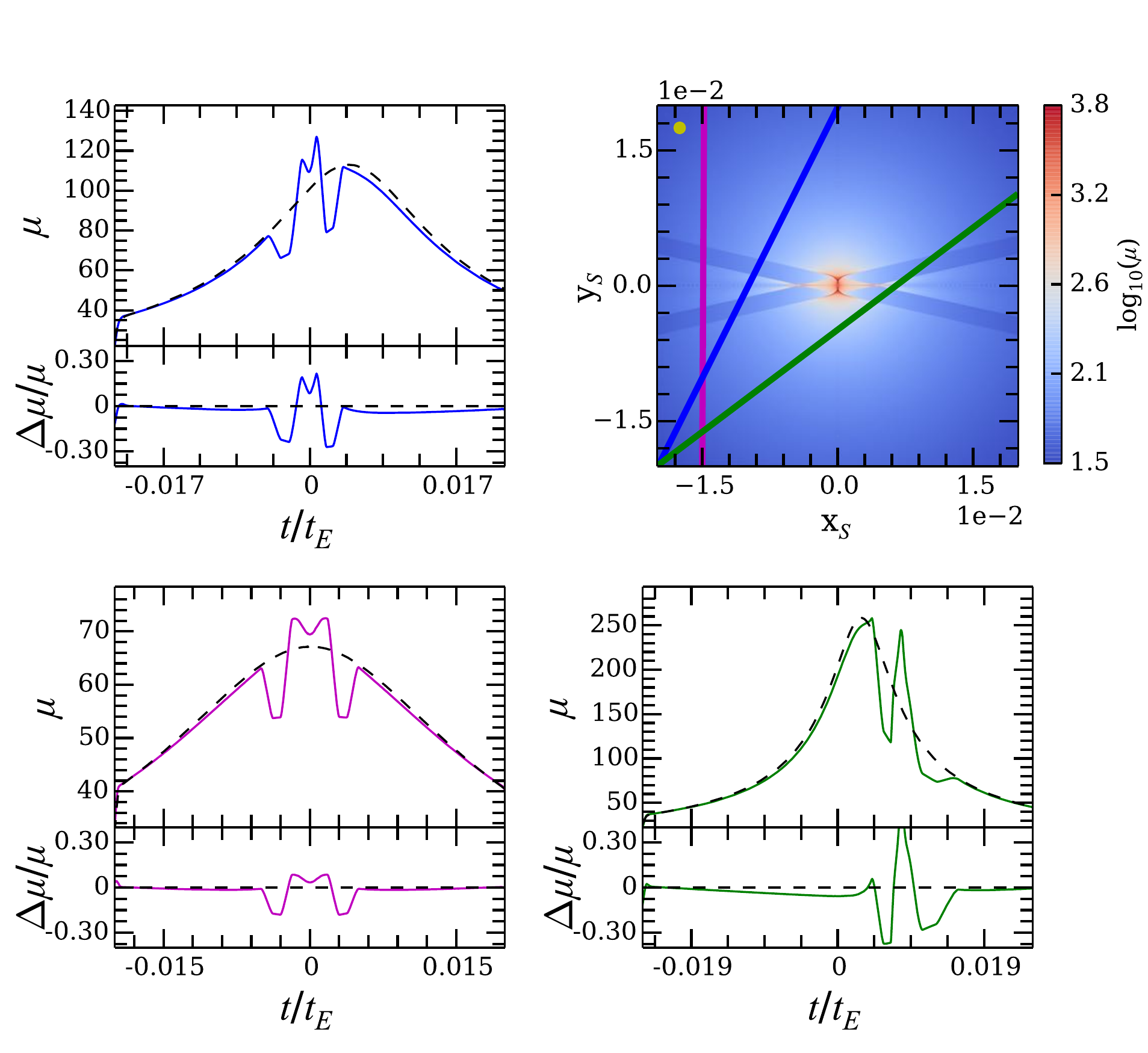}
\caption{\label{fig:a1to16q1e-3_lcurves}
Same as in Figure~\ref{fig:a12to14q1e-3_lcurves}, but with semi-major axes $a_i = 1$ and $a_o = 1.6$.
}
\end{figure}

\subsection{Pseudo-caustics}

Key to understanding the behavior of gravitational lenses is the study of the 
so-called ``caustic curves''. The caustic set contains the points in the source plane where the magnification $\mu$ for a point source is formally divergent and represents points where the determinant of the lens mapping Jacobian vanishes.
%(equation~\ref{eq:jac}) vanishes. 

Normally, the magnification is a smooth function for sources located away from the caustics. However, the belt+star lens models considered here possess regions in the source plane dubbed ``pseudo-caustics'', where the magnification is finite but changes discontinuously. Pseudo-caustics have been studied before in lensing systems with singular isothermal density distributions \citep[e.g.,][]{Kovner87,Wang97,Shin08,Rhie10,Lee14}, and they were associated with regions where the image multiplicity changes by one. In the systems considered here, the \pss\ can be associated either with regions where the image multiplicity changes or where an image of a source suddenly changes in size, depending on the width of the belt. They are found to possess a broad variety of morphologies across the parameter space of the models we consider. A more detailed analysis of the \pss\ is presented in \citet{Lake16}.

Since the boundaries of the \pss\ represent areas where the magnification map is non-differentiable (at least for the idealised constant-density belts considered here), they can in principle be found by finding points in the source plane where $|\nabla \mu| \rightarrow \infty$. 
In practice, we perform an inverse ray-shooting calculation \citep[e.g.,][]{Schneider86} by shooting rays from a uniform grid in the lens plane and collecting them on a grid in the source plane. The magnification of a cell in the source plane is then computed by dividing the number of rays it collects by the number it would collect in the absence of the lens. 

Figure~\ref{fig:panel_mag} shows magnification maps in the source plane for a 
belt with semi-major axes $a_i = 1.2,$ $a_o = 1.4$ and mass ratio $q = 10^{-3}$ across a range of inclinations. 
The \pss\ show up as the X-shaped features for the cases with high 
inclinations (e.g., $i \gtrsim 50^\circ$), which are fairly generic for the 
types of lenses we consider. The mass ratio $q = 10^{-3}$ may be rather 
unrealistic for asteroid belts, but the morphologies of the \pss\ do not change 
significantly when $q$ is lowered (the primary change being in the width of 
the pseudo-caustics), and so this case allows us to gain a qualitative understanding of the general behavior of belt+star lenses. When the inclination of the belt is low, the \pss\ are 
subdued. As shown in \citet{Lake16}, the deflection by a face-on belt is
zero for rays passing interior to the inner edge of the belt, while for those passing outside the belt it is identical to the deflection produced
by a point lens of dimensionless mass $q/(1+q)$ located at the origin. Therefore, for the case of a low mass ratio, the magnification map of the
belt+star system is close to that of the central star, meaning that such low-inclination belts do not cause significant perturbations to the star-only magnification pattern. 

As the inclination 
of the belt increases, the limbs of the X-shaped \ps\ feature fold down towards 
the $x_S$-axis and eventually merge when the belt is seen edge-on. For the 
edge-on case, the system exhibits some features degenerate with that of an 
equal-separation triple point-mass lens \citep[e.g.,][]{Danek15}, although it remains qualitatively different from the behavior of the magnification caused by edge-on circumstellar discs (see e.g. Figure 3 of \cite{Hundertmark09}, 
which considers spatially extended disks at a much larger distance from the central 
star than the asteroid belts considered here).

We find that the \pss\ are most pronounced for belts whose central semi-major axis $a_c = (a_i+a_o)/2$ is at or slightly beyond the Einstein ring radii 
($1 \lesssim a_c \lesssim 2$), as is the case of caustics for planetary 
microlensing \citep{Gould92}. 
%However, even focusing our attention only on the most favorable ring parameters still leaves a vast region of parameter space open for analysis.

When the width of the belt is increased, the \pss\ become smoother 
and ``smeared out'', although the effect is not too dramatic for moderate belt 
widths $\Delta a \lesssim 0.75$. For comparison, our own asteroid belt extends 
from $\sim$2.2 to $\sim$3.2~AU \citep{Petit01}, which if considered as a lens 
around a Sun-like star halfway between the Sun and the Galactic centre gives a belt width of $\Delta a \sim 0.25$.

Finally, we note that while Figure~\ref{fig:panel_mag} only shows a closeup of the magnification map near the centre of the source plane, the $X$-shaped \pss\ seen at large $i$ extend radially outward to cover a large region of the source plane. This means that source trajectories with comparatively large impact parameters ($b_0 \lesssim 0.5$) 
have a high likelihood of crossing a region of the \ps\ with significant perturbation to the source plane magnification, implying a high asteroid belt lensing 
probability. The \ps\ crossing leads to distinct features in the microlensing
light curves as shown in \S~\ref{sec:lcurves}.

\section{Model Light Curves}\label{sec:lcurves}

In this section, we present simulated light curves across a range of parameter 
combinations to demonstrate the expected behavior of belt+star microlensing 
events. Throughout this section, we assume the source star to be a
Sun-like star with mass of $1 {\rm M_\odot}$ and radius of $1 {\rm R_\odot}$ 
located near the Galactic centre with the belt+star 
lens located halfway to
the Galactic centre, i.e. we set $D_L = 4\kpc$ and $D_S = 8\kpc$. We take the 
lens star to be a Sun-like star of mass $M_{\rm star} = 1 \Msun$, which gives 
$\theta_E D_L \approx 4$AU and a dimensionless radius (radius in units of Einstein ring radius) for the source star of $\rho \sim 5.8 \times 10^{-4}$. 

\begin{figure}
\includegraphics[width=\columnwidth]{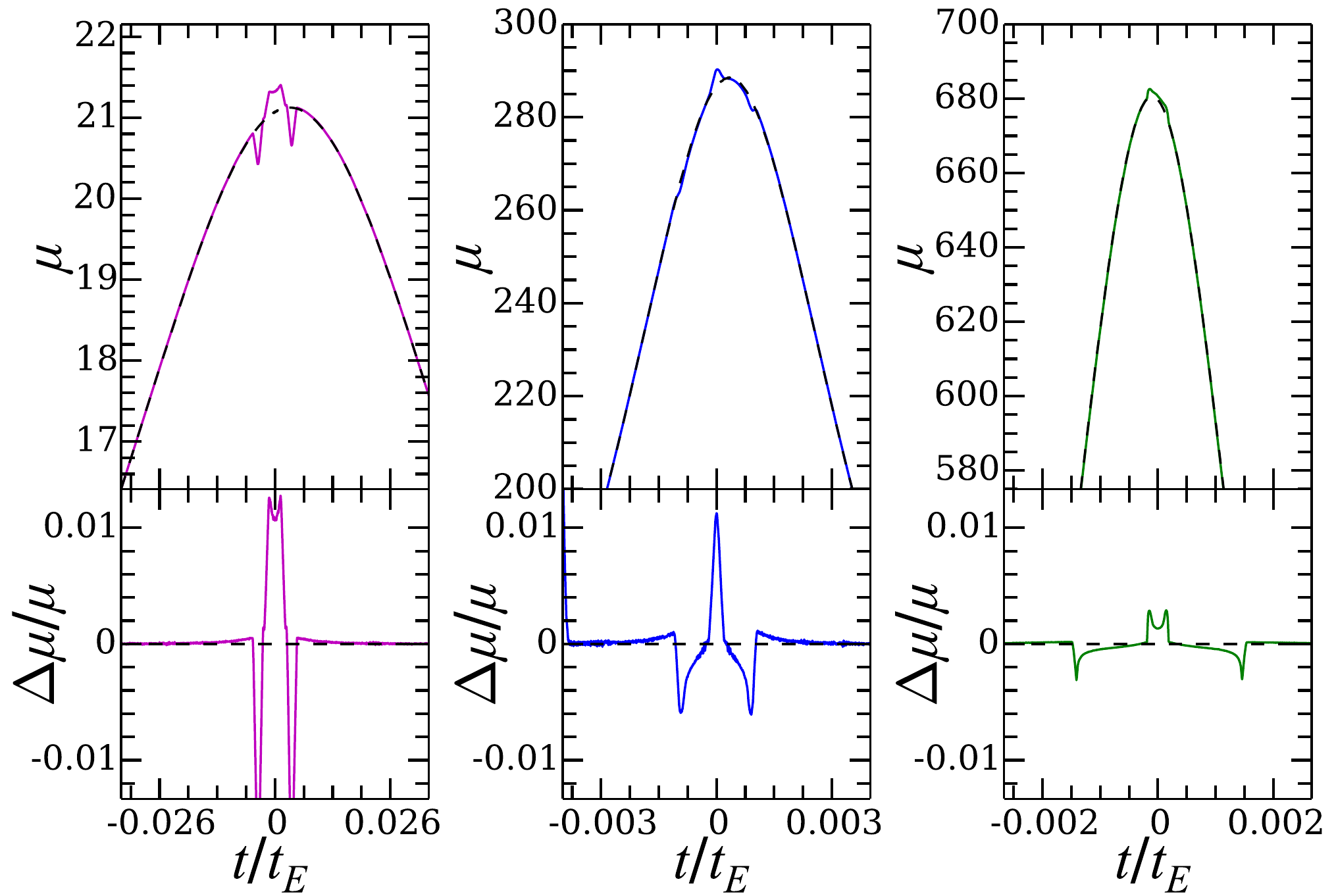}
\caption{\label{fig:3panel_lcurves}
Three sample microlensing light curves for a lens composed of a star and a belt with $q = 10^{-5}$, $i = 80^{\circ}$ and $\Delta a = 0.3$ for different semi-major axes and source trajectories. We assume a typical Sun-like source star located in the Galactic centre. From left to right, the belts are centred at $a = 1,~1.2,$ and $2$, respectively, and the trajectories of the source star have impact parameters $b_0$ = 0.05, 0.004 and 0.002, respectively. All trajectories make an angle of $\theta \sim 80^{\circ}$ with respect to the $x_S$ axis. 
}
%\vspace{20pt}
\end{figure}

\begin{figure}
\includegraphics[width=\columnwidth]{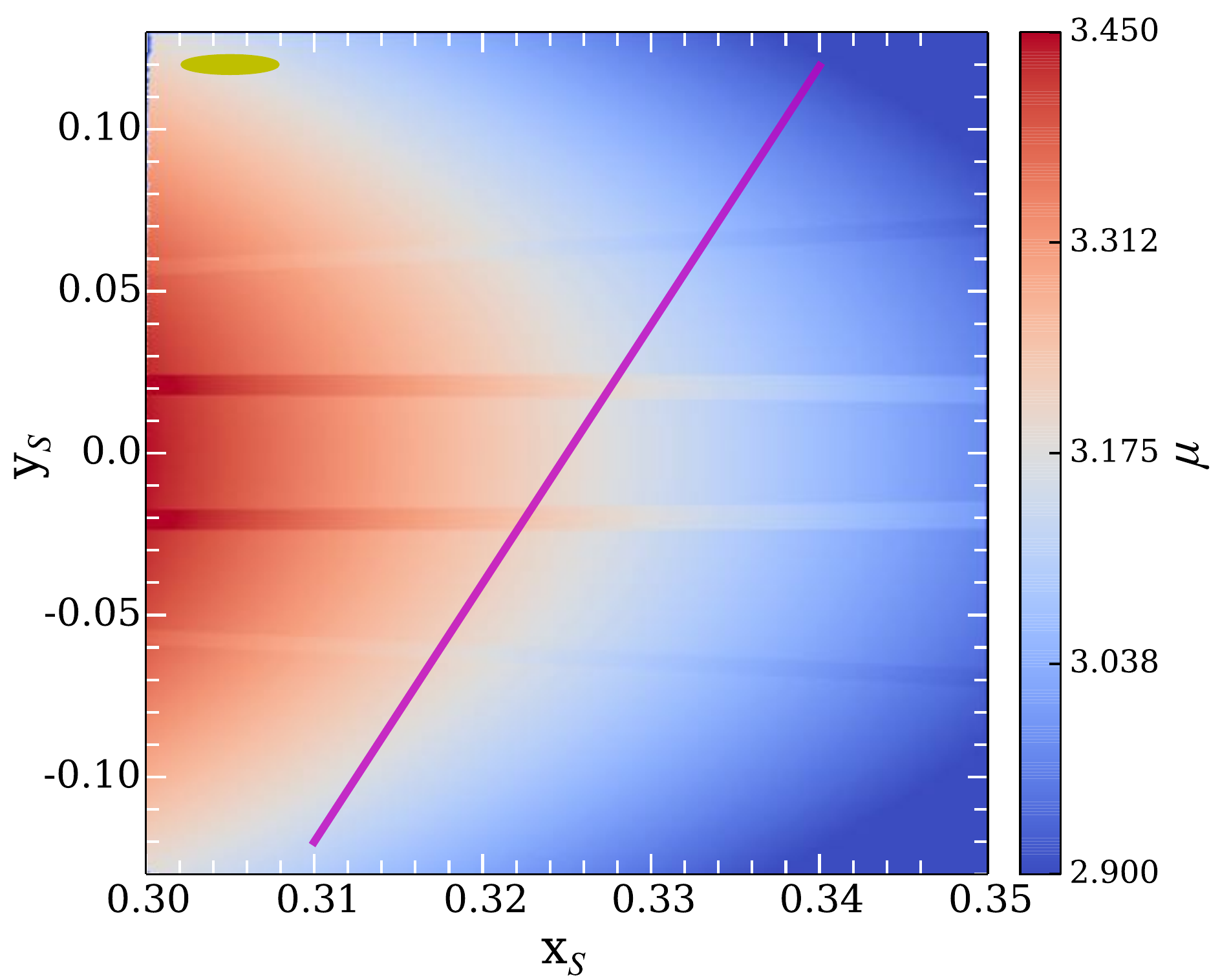}
\caption{\label{fig:large_map} Magnification map in the source plane for a region farther away from the centre for a star+belt lens with $q=10^{-5}$, $i=80^\circ$, $a_i=1.15$ and $a_o=1.45$. The horizontal stripe features are limbs of the X-shaped pseudo-caustic seen in Figures~\ref{fig:panel_mag} at high inclination. The pseudo-caustic features, showing up at large distances from the centre, implies a comparatively large probability of belt lensing events. The purple line shows a trajectory with impact parameter $b_0\sim$0.3, proceeding from bottom left to top right, which corresponds to the microlensing light curve shown in the left panel of Figure~\ref{fig:furtherout_3panel}. 
{ Note that in this plot the ranges along the 
$x_S$ and $y_S$ axes are different (about $1:5$). A circular source 
(corresponding to an $R = 5R_\odot$ source star) is illustrated by the 
yellow ellipse in the upper-right corner.}
}
\end{figure}

\begin{figure}
\includegraphics[width=\columnwidth]{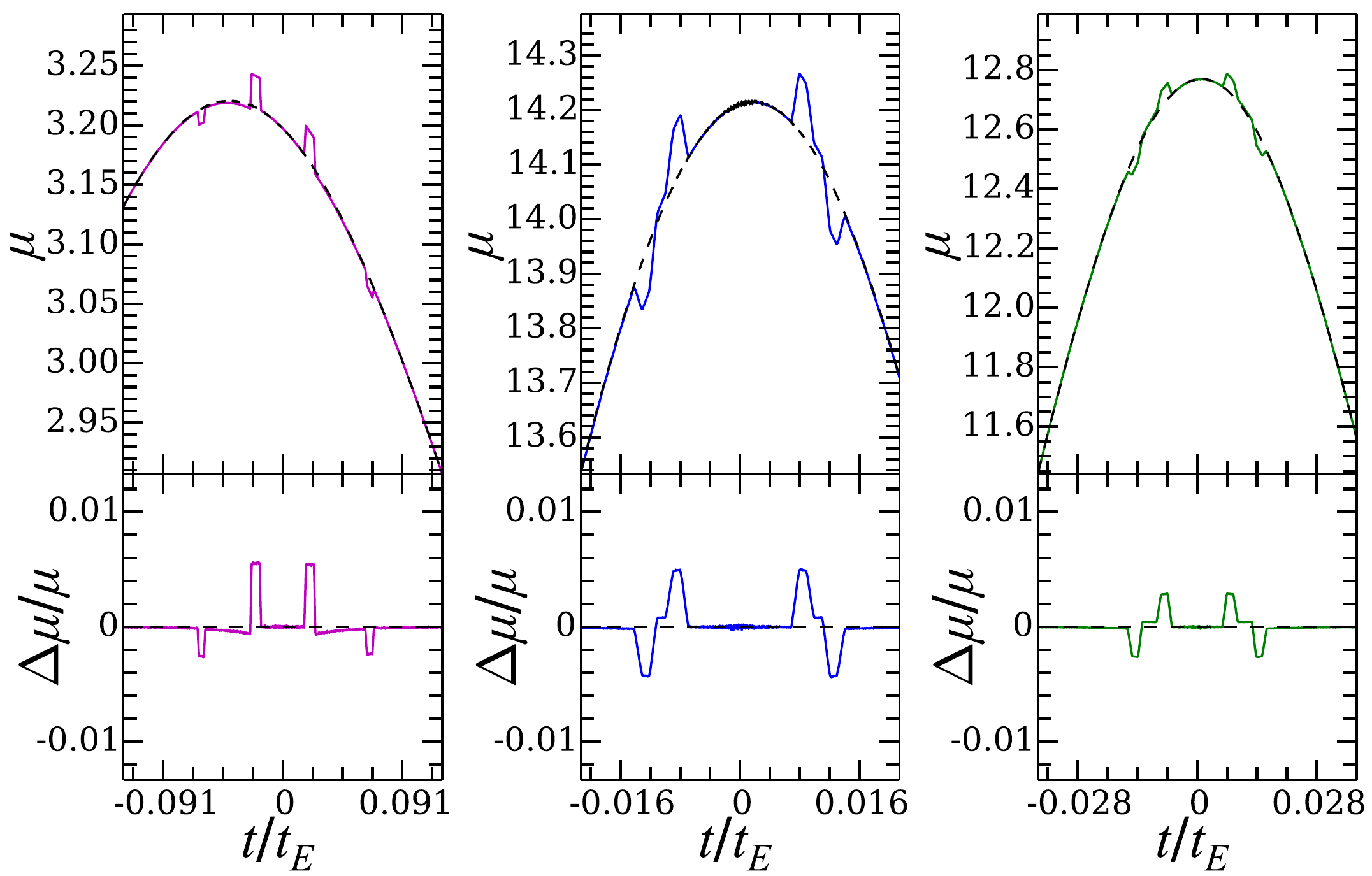}
\caption{\label{fig:furtherout_3panel}
Three more sample microlensing lightcurves for a lens composed of a central
star and a belt with $q = 10^{-5}$ and $i=80^\circ$. In the left, middle,
and right panel, the belt is centred at $a =1.3$, 1.3, and 1.6, respectively,
with the belt width set to $\Delta a=0.3$. From left to right, the trajectories of the source star have impact parameters of $b_0 \sim 0.3$, 0.08, and 0.1, with all trajectories making an angle of $\theta \sim 85^{\circ}$ with respect to the $x_S$ axis. These examples demonstrate that the large 
extent of the \pss\ allows them to leave signatures even in 
comparatively low-magnification events. 
}
\end{figure}

We first consider cases with belt-to-star mass ratio
of $q=10^{-3}$. While this is a rather high-mass (probably unrealistic)
asteroid belt, compared to the one in our solar system, it provides a clear
illustration for the main microlensing features. The cases with lower mass
ratios share similar features and will be presented afterward.
Figure~\ref{fig:a12to14q1e-3_lcurves} shows three model light curves for a belt+star system with $a_i = 1.2$, $a_o = 1.4$, $q=10^{-3}$, and $i = 80^{\circ}$. 
%$q=10^{-3}$ is an unrealistic mass ratio to be sure, but the qualitative features of the light curves shown here carry down to lower mass ratios, and so we stick with $q=10^{-3}$ for illustration. 
We focus on trajectories relatively close to the centre of the source plane, which lead to high magnification
microlensing events. 
The top-left panel shows the magnification map of the central region in the 
source plane, superimposed with the corresponding source trajectories for 
three light curves in the other three panels. The yellow circle at the top-left 
corner of the magnification map corresponds to the size of the source star, 
and we account for the finite source effect in calculating the light curves.
In each light curve, the black dashed line shows the light curve for the 
star-only lens, with the bottom subpanel showing the fractional deviation between the belt+star 
and star-only light curves. The time is given in units of $t_E$, the Einstein radius
crossing time. For a typical lens-source relative velocity of $v_{\perp} =110 ~\kms$, $t_E$ is 
$\sim$65 days. 

The distinct X-shaped \ps\ feature in the magnification map shows up in each 
of the three light curves. 
Given the shape of the \pss, the source
has a high likelihood of crossing the \pss\ twice, resulting in a light curve 
that exhibits a rapid decrease and increase in magnification, followed by a 
corresponding increase and decrease some time later. The 
maximum fractional change in magnification, $\Delta\mu/\mu$, can be as high
as $\sim$50\%. An observed light curve exhibiting these features would provide 
strong evidence for the presence of a belt+star lens. 

Figure~\ref{fig:a12to14_60_q1e-3_lcurves} is the same as Figure~\ref{fig:a12to14q1e-3_lcurves} but with the inclination reduced to $i=60^{\circ}$. Reducing the inclination has the effect of ``folding up'' 
the X-shaped \pss\ about the $y_S$-axis, and increasing the separation between 
the high and low magnification areas of the \pss. Additionally, the width of 
the \pss\ becomes smaller and the perturbations to the star-only light curve 
decrease.

Figure~\ref{fig:a07to09_q1e-3_lcurves} shows a case where both the inner and outer semi-major axes of the belt lie within the Einstein radius, with $a_i = 0.7$ and $a_o = 0.9$. Lowering the semi-major axes of the belt folds the 
X-shaped feature down onto the $x$-axis and smoothens the \pss. 
The maximum relative magnification $\Delta\mu/\mu$ between the belt+star case 
and the star-only case decreases to 5-15\%. In general, decreasing the 
semi-major axes even further results in a smaller perturbation to the 
star-only light curve. 

Figure~\ref{fig:a1to16q1e-3_lcurves} is again the same as 
Figure~\ref{fig:a12to14q1e-3_lcurves}, but with the thickness of the belt increased to $\Delta a = 0.6$, by setting $a_i = 1$, $a_o = 1.6$.
If a belt of this width were placed around a Sun-like lens star, a width of $\Delta a = 0.6$ would correspond to a physical width of $\sim2.5$AU. Thickening the ring smooths out the \pss, with the light curves retaining their overall shape but possessing slightly more subdued peaks and flatter troughs. 

The above $q=10^{-3}$ examples illustrate the expected features in the 
microlensing light curves for belt+star systems. More realistic systems
are likely to have much lower belt-to-star mass ratios. We show below that
the main features remain but with lower amplitudes. 
In Figure~\ref{fig:3panel_lcurves} we present three sample light 
curves for different trajectories for a system with a lower belt-to-star 
mass ratio, $q = 10^{-5}$, which corresponds to a belt mass of 
$\sim$$3M_\oplus$ for a Sun-like lens star. The inclination of the belt is
fixed at $i = 80^{\circ}$. From the left to right panel, the belts are 
centred at $a = 1$, 1.2, and 2, respectively, with width $\Delta a = 
0.3$, { and the source trajectories have impact parameters of 
$b_0 = 0.05$, 0.004 and 0.002, respectively. The trajectories are chosen 
to make an angle of $\theta \sim 80^\circ$ with respect to the $x_S$ axis,
ensuring that the sources cross the $X$-shaped \pss (similar to those seen
in the top-right panel of Fig.~\ref{fig:a12to14q1e-3_lcurves}) within
short time intervals. Trajectories of small impact parameters with smaller
$\theta$ angles would correspond to longer time intervals between \ps\ 
crossings, while trajectories of large impact parameters with smaller
$\theta$ angles are unlikely to experience strong \ps\ crossings.}
While the characteristic shape of the perturbations caused by the belt 
remains similar to the previous cases, the fractional change in 
magnification drops to the level of a percent. Approximately, the 
fractional perturbations with respect to the star-only light curve are  
proportional to the mass ratio $q$, of the order of $0.5(q/10^{-3})$.

The above light curves all possess small impact parameters, corresponding to high-magnification microlensing 
events. However, the pseudo-caustic features of the belt+star
lens can extend to large impact parameters, which will show up even in low-magnification events. Figure~
\ref{fig:large_map} shows a magnification map in the source plane for a region farther away from the centre. 
The belt has $q = 10^{-5}$ and is centred at $a=1.3$ with $\Delta a = 0.3$ and inclination $i = 80^{\circ}$. The 
horizontal stripe features are the limbs of the X-shaped pseudo-caustic seen earlier (e.g. in Fig.~
\ref{fig:panel_mag}). At large $r_S$ the high-$\mu$ and low-$\mu$ sections of the X-shaped pseudo-caustic separate from one another, which would make the peaks and troughs in the resulting light curves further apart. 
The fact that the pseudo-caustics extend out to large $r_S$ 
{ (e.g. beyond $r_S =0.35$ from the figure) implies a comparatively high 
probability for source trajectories to cross \pss. As a 
rather conservative estimate, by approximating the \pss\ as a linear feature 
along the $x_S$ axis, we find that among all trajectories with impact parameter
$b_0<0.35$, about 50\% of them cross the \pss\ at $x_S<0.35$. 
}

Finally, the light curve corresponding to the purple line (with impact parameter $b_0\sim$0.3) in Figure~
\ref{fig:large_map} is shown in the left panel of Figure~\ref{fig:furtherout_3panel}. The middle and right panels 
show two additional light curves for relatively low magnification events, by changing the belt to have
$a = 1.3$ and $1.6$ with $\Delta a = 0.3$ and impact parameter $b_0\sim$ 0.08 and 0.1, respectively. { As 
before, all the light curves are chosen to cross the $x_S$ axis, so that they intersect the \pss.} In all 
the three panels, the characteristic \ps\ crossing features caused by the belt clearly show up, with the relative 
magnification dropping to sub-percent levels. { The figure shows that 
the typical time scale for crossing a single \ps\ feature for the presented
light curves is about tenths of 
percent of the Einstein ring crossing time, i.e., on the order of hours for
microlensing events with Galactic centre sources and disk lenses, well within 
reach of future microlensing surveys like {\it WFIRST} (see more details in 
Section~\ref{sec:dis}).}

Overall, microlensing events from belt+star lenses can be generically 
characterized by a rapid decrease and increase in magnification followed by 
a reversed increase and decrease in magnification, caused by the perturbation 
from the asteroid belt. Such a semi-symmetric feature is seen 
across a wide range of belt semi-major axes, 
belt widths, belt inclinations, belt-to-star mass ratios, and source 
trajectories, making it a robust observational signal. Such a feature is not
usually seen in the light curves generated by other types of lensing systems,
except a moderate degeneracy with an equal-separation triple point-mass lens 
when the belt is seen nearly edge-on.
%Furthermore, it exhibits almost no degeneracy with the light curves generated by other types of lensing 
%systems (except a moderate degeneracy with an equal-separation triple point-mass lens when the belt is seen 
%very close to edge-on).

\begin{figure}
\includegraphics[width=.95\columnwidth,scale=.95]{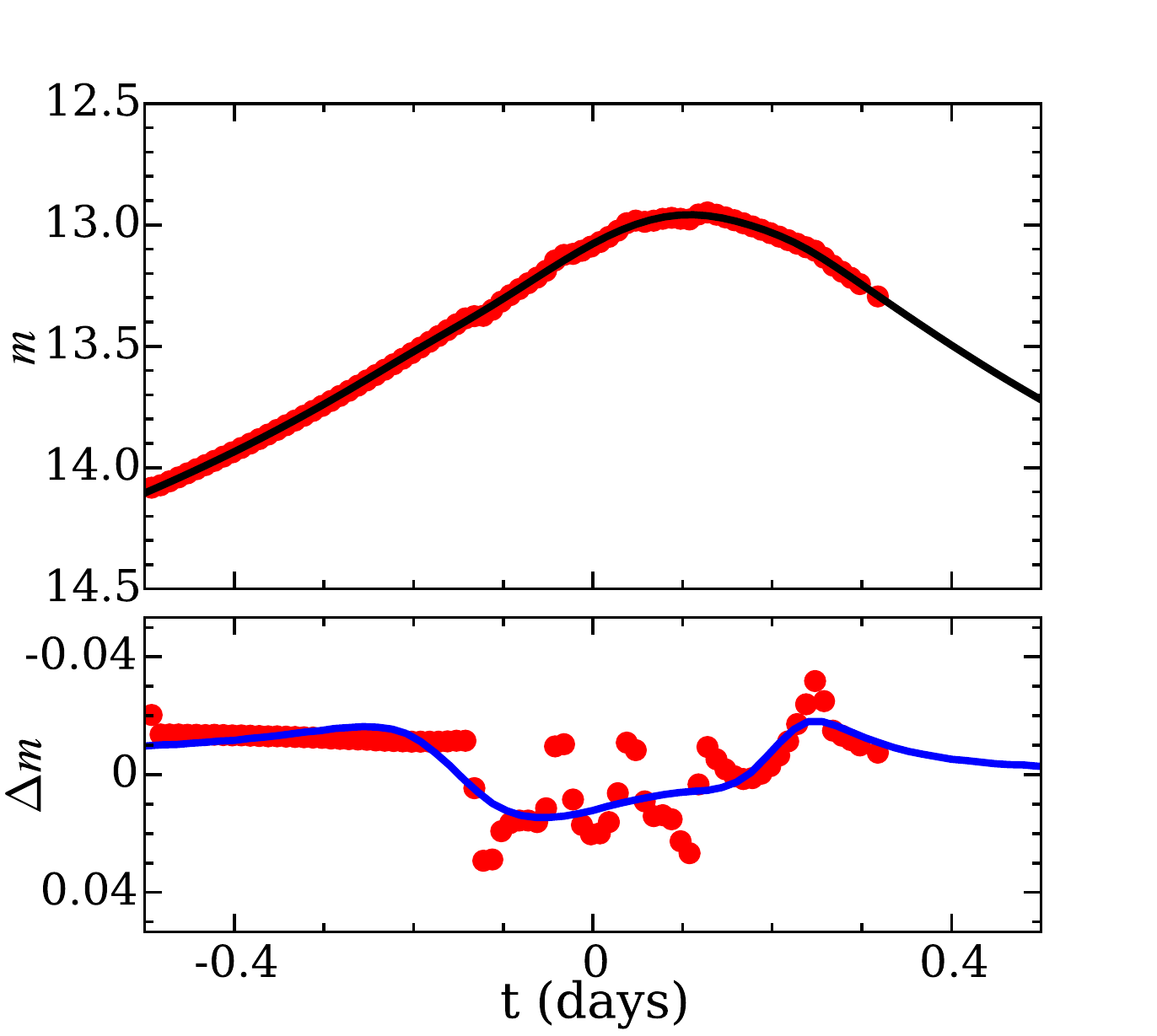}
\caption{\label{fig:test} An example of fitting a sample belt+star light curve with a planet+star 
lens model. The top panel shows the belt+star light curve (red dots) plotted against the best-fitting single lens model 
(black curve), shown in magnitudes. The bottom 
panel shows the difference between the belt+star light curve and the best-fitting single lens model (red dots) 
compared to the difference between the best-fitting planet+star lens model and the best-fitting single lens 
model (blue curve).  }
\end{figure}

\begin{figure}
\includegraphics[width=.97\columnwidth,height=210pt]{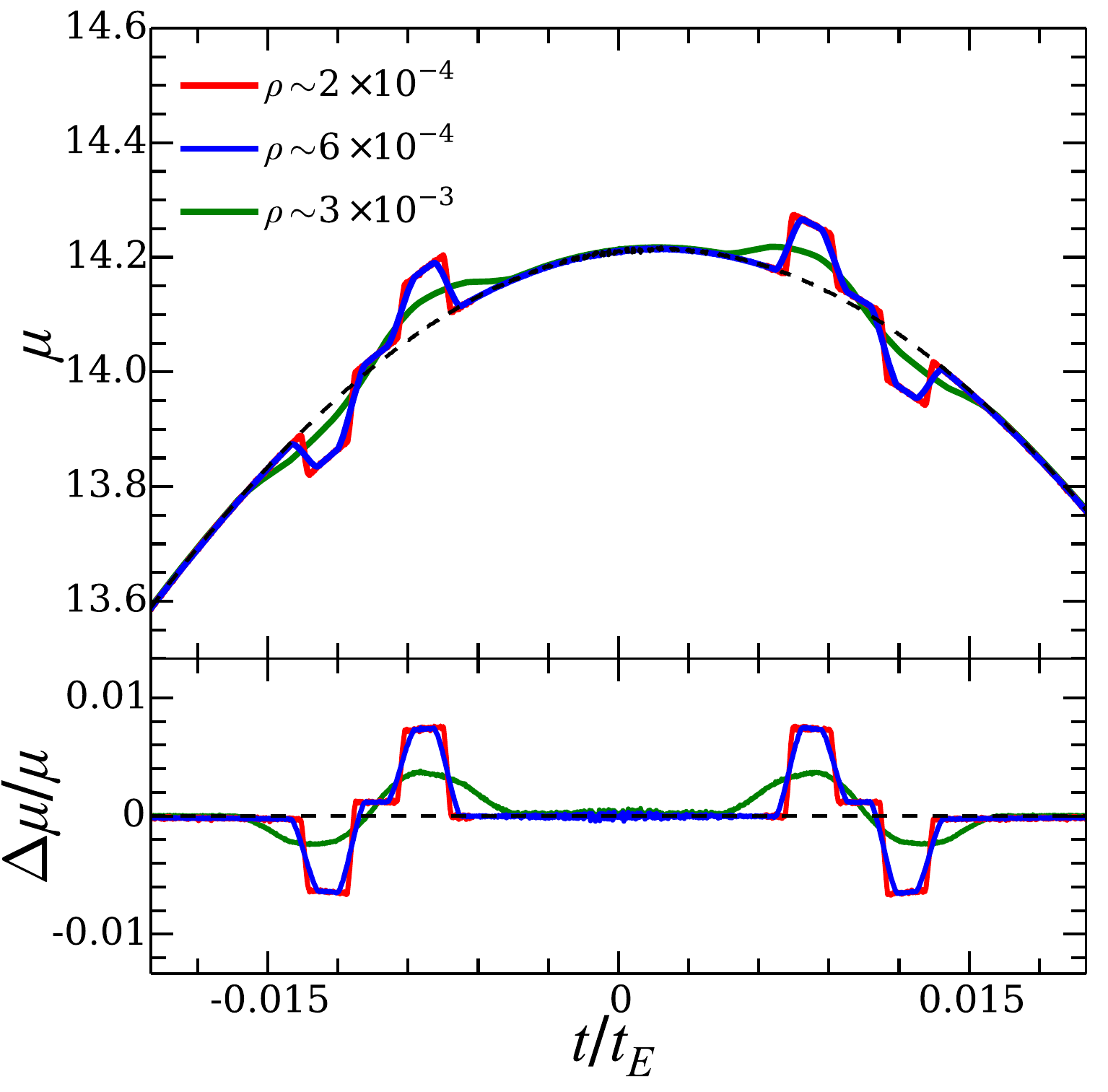}
\caption{\label{fig:finite_source_effect} The effects of changing the size of the source star for a sample lightcurve involving a belt with semi-major axes $a_i = 1.15,$ $a_o 
= 1.45$, mass ratio $q = 10^{-5}$, and $i = 80^{\circ}$. We assume a lensing geometry with $D_L = 4\kpc,$ $D_S = 8\kpc$. The dimensionless source radii $\rho$ 
correspond to an M-dwarf star with radius ${\rm 0.3R_\odot}$ (red curve), a Sun-like star with radius ${\rm 1 R_\odot}$ (blue curve), and a subgiant star with radius ${\rm 5 
R_\odot}$ (green curve). The dashed lines show the star-only case for comparison. } 
\end{figure}

The feature is distinct from that produced by planet microlensing.
Observationally, it would be unlikely for a belt+star microlensing event
to be misidentified as a planet+star one. To establish this quantitatively, we perform tests
by fitting belt+star microlensing curves with a planet+star lens model.
We produce theoretical belt+star microlensing light curve data and add
photometric errors of 5~mmag (reasonable for the best ground-based
microlensing surveys for high-magnification events). We then employ the
method in \citet{Dong09} to explore the parameter space of a planet+star
lens system to model the data. Figure~\ref{fig:test} shows an example, 
taken from a lens model with a belt of width $\Delta a = 0.2$ centered 
at $a = 1.25$ and with $i = 80^{\circ}$ and $q = 10^{-5}$.
In the top panel, the points are the densely sampled light curve of the
belt+star lens system, and the black curve is the best-fit from a single
lens model. The bottom panel shows the residuals to the best-fitting
single lens model as well as the difference between the best-fitting
planet+star model and the best-fitting single lens model (in blue).
The planet+star model tends to fit the broad deviations from the single-lens
model but fails to account for the small structures caused by the belt+star
lens. The $\chi^2$ difference between the planet+star and belt+star models
is 215.7 for 902 degrees of freedom (i.e., 5.1$\sigma$). For future applications,
a systematic investigation on the differences in planet+star, belt+star, and
planet+belt+star lens systems will be helpful.

The width of the \ps\ for systems with mass ratio $q=10^{-3}$ is roughly $10^{-3}$
times the Einstein ring radius { for source trajectories near the centre of the source plane, and it does not seem to vary substantially with the mass
ratio. The \pss\ typically become broader 
at a larger distance from the centre of the source plane. In addition, the 
width also depends on the inclination of the belts, with less inclined belts generically possessing thinner \pss.
}
For comparison, a Sun-like source star near the Galactic centre has a dimensionless
radius of $\rho \sim 5.8 \times 10^{-4}$.
This means that the magnification jump caused by a \ps\ crossing is smeared out by the finite
source effect, which ultimately limits the sensitivity to detect extrasolar
asteroid belt systems. 

Figure~\ref{fig:finite_source_effect} illustrates the effects of changing the size of the source star on the light curve plotted in the centre panel of Figure~\ref{fig:furtherout_3panel}.
%a belt+star lens event with $q = 5\times 10^{-5}$, $a_i = 0.9$, $a_o = 1.1$, and $i = 80^{\circ}$. 
As before, we assume a source star located in the galactic centre, with $D_L = 4\kpc$ and $D_S = 8\kpc$. For 
a typical ${\rm 0.3 M_\odot}$ M-dwarf source star with dimensionless radius $\rho \approx 1.7\times 10^{-4}$ 
(red curve), the perturbations caused by the belt are sharp, and are essentially the same as the curve created 
by an idealised point source. Changing the source to a Sun-like star with $\rho \approx 5.8 \times 10^{-4}$ 
(blue curve) results in a slight damping of the lightcurve, although the perturbations become significantly more 
damped when the source has a radius of $\rho \approx 2.9 \times 10^{-3}$, corresponding to either a massive 
or subgiant star with radius $5 R_{\odot}$ (green curve). 
{ 
Note that the typical width of the \pss\ here is about $1.5 \times 10^{-3}$.
The diameter of the M-dwarf source is much smaller than the \ps\ width, and the 
finite source has almost no effect (red curve). The Sun-like star source has 
a diameter comparable to (but still smaller than) the \ps\ width, and the 
finite source effect tends to 
slightly smooth the \ps\ features (blue curve). For the source with
radius of $5R_\odot$, the finite source effect causes the \ps\ features in the
light curve to be smeared significantly (green curve).
%much larger than the diameter of the source
%Note that the difference between the red and blue light curves is fairly minor, while 
%the difference between the blue and green light 
%curves is significant. This is because a dramatic change in the sharpness of the lightcurves occurs when the diameter 
%of the source star becomes larger than the width of the \ps\ features, so that the star's disk never lies entirely 
%within a \ps\ region (as is the case for the green curve). If the size of the source star is 
%already much smaller than width of the \pss, changing 
%its size slightly (e.g, changing between the red and blue curves) will have a small effect on the amount of smoothing in 
%the resulting light curve. 
%This also means that the finite source effect manifests itself differently for \ps\ crossings than for formal 
%caustic crossings: since the magnification transition for a source approaching a formal caustic is smooth and 
%the region of formally infinite magnification is infinitely thin, changing the source size will always affect the smoothness 
%of the resulting caustic-crossing light curves, even for small source sizes. 
Compared to the crossing of a formal caustic, where the magnification 
sharply increases to infinity, the finite source effect for the crossing of
a \ps\ with finite magnification jump is less dramatic. For smaller mass 
ratios ($q \sim 10^{-6}$), events of \ps\ crossing 
involving larger sources $(R\sim 5R_\odot)$ are unlikely to be observable except in the most extreme circumstances. }

\section{Summary and Discussion}\label{sec:dis}

We investigate detecting extrasolar asteroid belts around stars through
microlensing. Existence of asteroid belts leaves distinct signatures in 
microlensing light curves. Belt+star lenses generically create
so-called ``\pss'', regions in the source plane where the 
magnification of a source is finite but changes discontinuously. 
The probability of \ps\ crossing events is large for a wide range of 
belt configurations. In the light curves for most source trajectories,
the X-shaped \pss\ lead to a rapid fall and rise,
followed by a subsequent reversed rise and fall. Such a 
characteristic, semi-symmetric feature can be used to identify belt+star 
microlensing events.

%One common characteristic of the light curves generated by such a system is a rapid rise and fall in magnification, followed by a subsequent reversed fall and rise in magnification caused by the source passing over the \pss. If such a feature is observed in an actual light curve, it would provide very strong evidence for a belt + star lensing event.

The finite source effect means that sources at large distances 
(large values of $D_S$) are ideal, and that surveys should ideally target main 
sequence source stars rather than giants, as has been pointed out by 
\citet{Bennett96}. Detecting asteroid belts with $q\lesssim 5\times10^{-6}$ 
for Sun-like sources near the Galactic centre is an approximate lower bound for 
the masses of belts that are likely to be observed. The same type of
sources appear to be $\sim 8$ times smaller ($\rho \sim 7 \times 10^{-5}$) if located in the LMC. Therefore, targeting
stars in the LMC would enable us to detect lower mass asteroid belts, 
although most upcoming surveys (i.e. {\it WFIRST}; \citealt{Spergel15}) are 
being designed to target the Galactic bulge.

The small width of the \pss\ in the source plane also means that \ps\ crossing
events only last for a short period of time (e.g., minutes to hours for Galactic 
centre sources). The required high cadence rates for detection are similar 
to those in current surveys like MOA and OGLE \citep{Sumi10,Udalski15}. 
As a concrete
example for future surveys, we consider the photometric survey towards the
Galactic bulge by the {\it WFIRST} satellite. With a 52s exposure, the photometric
precision in a broad $H$ band is about $\sigma(H)\sim 10^{(2/15)(H-15)}$ 
mmag \citep{Gould15}, where $H$ is the apparent magnitude of the source. 
A ${\rm 0.3 M_\odot}$ M-dwarf source star (with absolute magnitude $M_H\sim 7.05$ mag; 
\citealt{Henry93}) in the Galactic centre has an apparent magnitude 
$H\sim 21.57$ mag, and the corresponding photometric precision is
$\sim 7.5$ mmag. So even for microlensing events with moderate magnifications, 
percent level features caused by extrasolar asteroid belts in the light 
curves can be detected with high significance.

%which is similar
%this necessitates having high cadence rates (on the order of tens of minutes) in order to capture the deviations to the light curve caused by the belt. This is not too restrictive, and is similar to the rates used by surveys like MOA and OGLE \citep{Sumi10}.

While microlensing features by extrasolar asteroid belts can be detected at
high precision, one may wonder whether or not asteroid belts similar to the idealised ones considered here actually exist. Clearly the key parameter here
is the mass ratio $q$ of the asteroid belt to its parent star.
%the 
%probability for ongoing and forthcoming surveys depends on the mass ratios of 
%asteroid belts to their parent stars. 
The present mass of our own asteroid 
belt is only $\sim5 \times 10^{-4} M_\oplus$ \citep{Petit01}, giving 
$q\sim 5\times10^{-9}$ even for a ${\rm 0.3M_\odot}$ M-dwarf parent star. 
Observing such a low-mass belt would be very unlikely, even for a survey 
targeting the LMC to reduce the finite source effect. However, our 
asteroid belt is thought to have been several orders of magnitude more massive 
in the early years of our solar system, with a primordial mass of 
$\sim1 M_\oplus$ before it lost most of its mass to ejection by interactions 
with other planets \citep{Weidenschilling77, Petit01} over a time scale of
a few Myr. For systems that lack massive planets orbiting near the belt, the 
lifetime of early, massive asteroid belts may be significantly longer. 
Furthermore, the {\it Kepler} mission has found evidence that extrasolar super 
Earths are likely common \citep[e.g.][]{Howard12,Dong13}, and that the extrasolar 
minimum mass solar nebular (MMSN) can have surface densities many times 
higher than the solar MMSN \citep{Chiang13,Raymond14}. Like extrasolar planets, 
extrasolar asteroid belts may be very diverse, and there may exist 
extrasolar asteroid belts that are orders of magnitude more massive than
the solar system's asteroid belt. 

Cold planetesimal discs / asteroid belts of masses on the order of a few 
$M_\oplus$ at a distance a few AU from a solar mass star can be dynamically 
stable over Gyr timescales \citep{Heng10}.  
There are hints from observations that massive asteroid belts or ring-like
structures exist. For example, the inferred parent mass of the asteroid belt 
around the A-type star $\zeta$ Leporis from the observation of dust emission a 
few AU from the star is about $0.1M_\oplus$ \citep{Chen01,Moerchen07b}, substantially more 
massive than our own. Another example comes from the eclipse light curve 
of J1407 \citep{Kenworthy15}, which can be interpreted as caused by a 
companion star, J1407b, with a giant ring system ($\sim 1 M_\oplus$, 
extending to a radius of $\sim 0.6$AU). 

%If belts with masses comparable to $\mathcal{O}(10)$ times the mass of the moon (giving $q\sim2\times10^{-6}$ for typical m-dwarf host star) exist, they should in principle be detectable in microlensing events, provided a reasonably favorable inclination ($i \gtrsim 75^{\circ}$) and a semi-major axis of $a \gtrsim 1$, particularly if the lensing survey targets the LMC. One may naturally wonder wether or not such structures would be dynamically stable on their own, independent of their interactions with planetary companions. While we treat the belts in our analysis as uniform, in reality they are made out of many interacting masses. We may safely assume the belts we consider to be ``cold'', i.e., that collisions between asteroids never occur. With this assumption, planetesimal discs / asteroid belts of masses on the order of a few $M_{\bigoplus}$ can be dynamically stable over Gyr timescales, provided their semi major axes are sufficiently large (on the order of a few AU) \citep{Heng10}. Additionally, the recent discovery of a very massive $(M_{belt}\sim1M_{\bigoplus},~q\sim10^{-5})$ ring system around the star J1407B \citep{Kenworthy15} implies that massive asteroid belts may not be prohibitively uncommon.

Overall, it is not unlikely that belt structures with masses of 
$\mathcal{O}(10)$ times the mass of the moon (corresponding 
$q\gtrsim 2\times10^{-6}$ for a typical M-dwarf host star) exist, which
in principle can produce detectable microlensing features. The favorable 
configurations include a reasonably high inclination ($i \gtrsim 70^{\circ}$), 
a central semi-major axis of $1 \lesssim a_c \lesssim 2$, and a belt width of $\Delta a \lesssim 0.75$. 

It is natural to assume that most 
asteroid belts will have semi-major axes close to the snow line, beyond which 
the temperature from their parent stars is low enough to allow ices to form. 
This is encouraging, given that the snow line 
$r_{\rm snow} \approx 2.7 (M/\Msun)^{1/3}$~AU \citep{Martin13} is fairly close 
to $\theta_E D_L$ (around which the perturbations caused by the belt are most 
significant; see equation~(\ref{eq:xi0})) for a typical Galactic microlensing 
event. Other 
disk-like objects like Kuiper belt analogues will be more 
difficult to detect observationally, since they will typically be located 
many Einstein radii away from their host stars. Therefore, we focus on asteroid belts in this study.

In our calculation of the microlensing signal, we assume a smooth asteroid belt with constant mass density, 
as this idealisation allows us to gain a simple understanding of how asteroid belts behave as gravitational 
lenses. In reality, asteroid belts are made of discrete bodies, with a smooth belt corresponding to the limit of 
small bodies. Depending on the mass
distribution of the belt, effects arising from the discrete nature of the true mass distribution need to be 
investigated and  
need to be accounted for in modeling the microlensing light curve in 
observations. We note that a study of microlensing caused by planetesimal discs modeled as an ensemble 
of point masses was considered in \cite{Heng09}. 

All in all, it does not seem unreasonable to consider the existence of 
long-lived massive asteroid belts. If they exist, microlensing is an ideal 
method for observing them. Observational techniques are continuing to improve, 
with microlensing surveys reaching increasingly high levels of sensitivity. 
New space-based surveys like {\it WFIRST} will be able to detect 
Mercury-mass planets \citep{Spergel13} and will be well-poised to make the 
first detections of extrasolar asteroid belts, which will provide useful 
information on the formation of planets and dynamical evolution of asteroid 
belts.

\section*{ACKNOWLEDGEMENTS}
We thank the anonymous referee for helpful comments.
The support and resources from the centre for High
Performance Computing at the University of Utah are gratefully acknowledged.
S.D. is supported by the Strategic Priority Research Program ``The
Emergence of Cosmological Structures of the Chinese Academy of
Sciences'' (Grant No. XDB09000000) and Project 11573003 supported by
NSFC.

\bibliographystyle{mnras}
\bibliography{bibliography}

\end{document}